\documentclass[letterpaper,aps,preprint,nofootinbib]{revtex4}%
\usepackage{amsfonts}
\usepackage[utf8]{inputenc}
\usepackage{amsmath}
\usepackage{amssymb}
\usepackage{float}
\usepackage{graphicx}%
\providecommand{\U}[1]{\protect\rule{.1in}{.1in}}

\begin{document}
\title{Stabilizing Homogeneous Black Strings in AdS}
\author{Adolfo Cisterna$^{1}$, Carla Henr\'iquez-B\'aez$^{2}$, Julio Oliva$^{2}$}
\affiliation{$^{1}$ Vicerrector\'ia Acad\'emica, Toesca 1783, Universidad Central de Chile, Santiago, Chile}
\affiliation{$^{2}$ Departamento de F\'isica, Universidad de Concepci\'on,
Casilla 160-C, Concepci\'on, Chile.}

\begin{abstract}
It is known that homogenous black strings in flat spacetime are unstable under long wavelength perturbations. Recently, by including scalar fields that depend on the extended coordinates only, homogenous black strings have been constructed in the presence of a negative cosmological constant, therefore implying that generic perturbations will contain both metric and scalar fluctuations. Here we show that for such generic perturbations homogeneous black strings with a negative cosmological constant are perturbatively stable at linear level, regardless the size of the Schwarzschild-AdS black hole located on the brane.
\end{abstract}
\maketitle

\section{Introduction} In General Relativity (GR) in vacuum, one can trivially oxidate black holes to
create homogeneous black strings and black $p$-branes in higher dimensions.
These objects are an explicit example of the new phenomena that higher-dimensional gravity offers \cite{Obers:2008pj}. Indeed, they express a very rich dynamics since they were proved to suffer from the so-called Gregory-Laflamme (GL) instability \cite{Gregory:1993vy}, a long-wavelength perturbative instability triggered by a mode that travels along the extended direction of the horizon, 
mimicking the spectrum of the Rayleigh-Plateau instability in fluid dynamics \cite{Cardoso:2006ks}. 
Beyond the validity of the linear regime, there is enough evidence that for long
times, the non-linear effects of GR trigger a violation of cosmic censorship,
since the black string horizon will pinch-off in a finite amount of time as seen by an
asymptotic observer \cite{Choptuik:2003qd, Lehner:2010pn}. Such instability pervades asymptotically
flat, black objects in higher dimensions, leading to a similar pinch-off of
the rapidly rotating Black Ring horizon \cite{Figueras:2015hkb} as well as of the rapidly
rotating Myers-Perry black hole in dimensions $D\geq6$ \cite{Figueras:2017zwa}.
It is known that the inclusion of a negative cosmological constant spoils any direct, homogeneous
oxidation of the asymptotically AdS black holes (see e.g. \cite{Mann:2006yi}), nevertheless, as recently
reported in \cite{Cisterna:2017qrb} this obstruction can be circumvented in five dimensions by the presence
of a massless scalar $\psi\left(  y\right)  $ such that $d\psi\left(
y\right)  =\lambda dy$, where the integration constant $\lambda$ gets fixed in terms of the bared
cosmological constant, and $y$ stands for the coordinate along the extended direction\footnote{These configurations can be though of as Freund-Rubin background \cite{Freund:1980xh} for a fundamental $0$-form, with a $1$-form field strength $d\psi$ that is proportional to the volume form of the extended direction.}. This construction can be extended to flat $p-$branes by
the introduction of $p$ massless scalars $\psi_{i}\left(  y^i\right)  $ with
$i=1,...,p$ such that%
\begin{equation}
d\psi_{1}\wedge...\wedge d\psi_{p}\sim dy^{1}\wedge...\wedge dy^{p}\ ,
\end{equation}
with $y^{i}$ the Cartesian coordinates of the extended flat directions. The
purpose of the present work is to prove the linear, mode stability of these
configurations under generic perturbations. These perturbations lead to an instability in the asymptotically flat case,
nevertheless, in the presence of a cosmological constant they do not trigger
any instability at all.

\section{Homogeneous Black Strings in AdS} Homogenous Black Strings and $p$-branes are constructed out from $D=d+p$ dimensional spacetimes of the form
\begin{equation}
\mathring{g}_{AB}dx^Adx^B=g_{\mu\nu}dx^{\mu}dx^{\nu}+\delta_{ij}dy^idy^j
\end{equation}
where the $D-$dimensional metric $\mathring{g}_{AB}$ is the direct product of a $d-$dimensional black hole metric $g_{\mu\nu}$ and a $p-$dimensional flat Riemannian metric $\delta_{ij}$.
To circumvent the evident incompatibility of the equations of motion induced by the presence of a negative cosmological constant, we use $p$ minimally coupled scalar fields $\psi_i=\psi_i(y^i)$, that dress the extended coordinates in such a manner that no incompatibility arises \cite{Cisterna:2017qrb}.
Then, the Einstein-Klein-Gordon system%
\begin{align}
G_{AB}+\Lambda g_{AB}  & =\sum_i^p\frac{\kappa}{2}\left(  \partial_{A}\psi_i\partial
_{B}\psi_i-\frac{1}{2}g_{AB}(\partial\psi_i)^2\right)  \ ,\\
\square\psi_i & =0\ ,
\end{align}
admits the following, homogeneous black string solution%
\begin{align}
ds^{2}&=\mathring{g}_{AB}dx^{A}dx^{B}\\
&=-f(r)dt^{2}+\frac{1}{f(r)}dr^{2}%
+r^{2}d\sigma^2_{d-2}  +\delta_{ij}dy^idy^j%
\ ,\label{backgroundmetric}%
\end{align}
with $d\sigma_{d-2}$ the line element of a $(d-2)$-dimensional hypersphere and 
\begin{align} \label{lapse}
f(r)&=1-\frac{2m}{r}-\frac{2\Lambda}{(d-1)(d+p-2)}r^{2}\\ 
\lambda^{2}&=-\frac{4\Lambda}{\kappa(d+p-2)}\\
\psi_i&=\lambda y^i\ . \label{backpsi}  
\end{align}
Clearly, the bared cosmological constant $\Lambda$ has to be negative, and for
further simplicity, and without losing generality we will fix $\Lambda$ in such a way that for the corresponding spacetime dimension, $\lambda=1$.\\
The next section is devoted to prove the stability, under generic linear perturbations of these configurations. Although the computation can be carried out in arbitrary dimension, leading to the same conclusions, for the sake of concreteness we focus on the five-dimensional case, even though we provide some comments on the arbitrary $D$ case in the conclusions.

\section{Linear perturbations} The five-dimensional AdS black string is given by the configuration \eqref{backgroundmetric} with
\begin{align}
f(r)&=-\frac{2\Lambda r^2}{9}-\frac{2m}{r}+1 \hspace{0.2cm},\ \lambda^2=-\frac{4\Lambda}{3\kappa}\\
\psi&=\lambda y
\end{align}
on which we fix $\kappa=2$ and $\Lambda=3/2$ such that $\lambda=1$.
A generic, spherically symmetric perturbation of this metric will contain both, a fluctuation
of the scalar field as well as a fluctuation of the metric, therefore, as for the
asymptotically flat black string we introduce the following metric perturbation
$g_{AB
}\rightarrow\mathring{g}_{AB}+\epsilon h_{AB}$ with
\begin{equation}
h_{AB}(r)=e^{\Omega t+iky}%
\begin{pmatrix}
h_{1}(r) & h_{2}(r) & 0 & h_{6}\left(  r\right)  \\
h_{2}(r) & h_{3}(r) & 0 & h_{7}\left(  r\right)  \\
0 & 0 & h_{4}(r)\sigma_{2} & 0\\
h_{6}\left(  r\right)   & h_{7}\left(  r\right)   & 0 & h_{5}\left(  r\right)
\end{pmatrix}
\ ,
\end{equation}
while for the scalar field we consider
\begin{equation}
\psi\rightarrow\mathring{\psi}+\epsilon\chi=y+\epsilon\Phi(r)e^{\Omega
t+iky}\ .\label{scalarpert}%
\end{equation}
Clearly the metric and scalar field fluctuations preserve the spherical symmetry of the background. $\epsilon$ is the perturbative parameter. If we are able to find
bounded solutions of the linearized system, for positive values of
$\Omega$, we would have found an unstable mode, while the absence of such modes will imply the linear stability of our configuration. The linearized Einstein-Klein-Gordon system leads to%
\begin{align}\nonumber
&-\frac{1}{2}\mathring\triangle_Lh_{AB}+\frac{1}{2}\mathring{g}_{AB}h^{CD}\mathring{R}_{CD}+\frac{1}{4}\mathring{g}_{AB}\mathring{g}^{CD}\mathring{\triangle}_Lh_{CD}+h_{AB}\left(\Lambda-\frac{1}{2}\mathring{R}\right)=T^{(1)}_{AB} \ ,
\end{align}
\begin{align}
&\mathring\Box\chi-\mathring\nabla_Ah^{AB}\mathring\nabla_B\mathring\psi-h^{AB}\mathring\nabla_A\mathring\nabla_B\mathring\psi+\frac{1}{2}\mathring\nabla_Ah\mathring\nabla^A\mathring\psi=0\label{fieldfluc} \ ,
\end{align}
where $\mathring\triangle_L$ is the Lichnerowicz operator of the background  spacetime, and
\begin{align}
T^{(1)}_{AB}&=\frac{1}{2}(\partial_A\mathring\psi\partial_B\chi+\partial_A\chi\partial_B\mathring\psi)-\frac{1}{4}\mathring g_{AB}\mathring g^{CD}(\partial_C\psi\partial_D\chi+\partial_C\chi\partial_D\mathring\psi)\\ \nonumber
&-\frac{1}{4}\partial_C\mathring\psi\partial_D\mathring\psi(h_{AB}\mathring g^{CD}-\mathring g_{AB}h^{CD}) \ . \nonumber
\end{align}
In the presence of matter in GR, it is consistent to impose the following gauge condition%
\begin{equation}\label{gaugecond}
\mathring{\nabla}_{A}h^{AB}=\frac{1}{2}\mathring{\nabla}^{B}h_{\ C}^{C}\ ,
\end{equation}
which reduces the equation for the
scalar field fluctuation (\ref{fieldfluc}) to%
\begin{equation}
\mathring{\square}\chi-h^{AB}\mathring{\nabla}_{A}\mathring{\nabla}%
_{B}\mathring{\psi}=0\ .
\end{equation}
Even more, the second term vanishes since for the background metric
(\ref{backgroundmetric}), the components $\mathring{\Gamma}_{AB}^{y}$ of the Christoffel symbol identically vanish, and we are therefore left with the Klein-Gordon equation
on the background metric for the scalar field fluctuation. Introducing the
separation (\ref{scalarpert}) one is left with the following ODE%
\begin{align}
(r-r_{+})^{2}(r^{2}+r_{+}r+r_{+}^{2}+3)r\frac{d^{2}\Phi}{dr^{2}}&+&\nonumber\\
(r-r_{+})(4r^{3}-r_{+}^{3}+6r-3r_{+})(r^{2}&+r_{+}r+r_{+}^{2}+3)\frac{d\Phi
}{dr}-3r^{2}((r-r_{+})k^{2}+3r\Omega^{2})\Phi=0\ .\label{ecur}%
\end{align}
The regular asymptotic behaviors provided by this equation, near the
horizon and infinity yield%
\begin{align}
\Phi\left(  r\right)    & \sim\left(  r-r_{+}\right)  ^{\frac{r_{+}\Omega
}{r_{+}^{2}+1}}\left(  1+\mathcal{O}(r-r_{+})\right)  \text{ as }r\rightarrow
r_{+}\ ,\\
\Phi\left(  r\right)    & \sim r^{-\frac{3}{2}-\frac{\sqrt{12k^{2}+9}}{2}%
}\left(  1+\mathcal{O}(r^{-1})\right)  \text{ as }r\rightarrow+\infty\ .
\end{align}
For positive values of $\Omega$, connecting these two asymptotic behaviors is
impossible. This is proved as follows. Any bounded solution of
(\ref{ecur}) has to vanish both at infinity and at the horizon, therefore there must exist a
turning point in the domain of outer communication. At such point  $r=r_{\ast}>r_{+}$ one must have%
\begin{equation}
\frac{1}{\Phi\left(  r_{\ast}\right)  }\frac{d^{2}\Phi\left(  r_{\ast}\right)
}{dr^{2}}=\frac{3r_{\ast}((r_{\ast}-r_{+})k^{2}+3r_{\ast}\Omega^{2})}%
{(r_{\ast}-r_{+})^{2}(r_{\ast}^{2}+r_{+}r_{\ast}+r_{+}^{2}+3)}\ ,%
\end{equation}
and since the rhs is strictly positive, no such turning point exists. This
standard argument proves the non-existence of exponentially growing modes in
time, for generic perturbations of the black string, and therefore its stability!

\section{Further remark} We have proved that the same perturbation that leads to the GL instability of
asymptotically flat black strings, is not able to trigger an instability of
the homogeneous asymptotically $AdS_{4}\times R$ black strings of GR in five
dimensions, supported by a massless scalar. The presence of the scalar field
therefore stabilizes the black string solution. The results presented above
can be extended to arbitrary dimension $D=d+p$, and even beyond spherical symmetry. Indeed, using
the same gauge condition \eqref{gaugecond}, each of the perturbations of the $p$ scalar fields in equation \eqref{backpsi},
$\delta\psi_{\left(  i\right)  }=\chi_{\left(  i\right)  }$, will solve a
Klein-Gordon equation on the background metric \eqref{backgroundmetric}.
Introducing the separation $\chi_{\left(  i\right)  }=e^{\Omega t+i\vec
{k}\cdot\vec{y}}Y_{l}\left(  \sigma_{d-2}\right)  \Phi_{\left(  i\right)
}\left(  r\right)  $, leads to the ODE%
\begin{align}
r^{2}f^{2}\Phi_{\left(  i\right)  }^{\prime\prime}\left(  r\right)  +rf\left(
\left(  d-2\right)  f+rf^{\prime}\right)  \Phi_{\left(  i\right)  }^{\prime
}\left(  r\right) -\left(  \Omega^{2}r^{2}+l\left(  l+d-1\right)  f+\vec
{k}^{2}r^{2}f\right)  \Phi_{\left(  i\right)  }\left(  r\right)
=0\ ,\label{ardode}%
\end{align}
with $i=1,\ldots,p$ and where $Y_{l}\left(  \sigma_{d-2}\right)  $ stands for a spherical harmonics on
the $\sigma_{d-2}-$sphere. As usual due to the spherically symmetric nature
of the background, only the $l$ label of the harmonics appears. The relevant asymptotic behaviours are $\Phi_{\left(
i\right)  }\sim\left(  r-r_{+}\right)  ^{\Omega/f^{\prime}\left(
r_{+}\right)  }$ and $\Phi_{\left(  i\right)  }\sim r^{-\Delta_{+}}$, with
$2\Delta_{+}:=\left(  d-1\right)  +\sqrt{\left(  d-1\right)  ^{2}+4L^{2}%
\vec{k}^{2}}$ at the horizon and at infinity, respectively and $L$ is the
effective $AdS$ radius of the asymptotic region of the black hole on the
brane. As before, equation (\ref{ardode}) does not admit regular solutions for
positive $\Omega$ and therefore, the system is stable.

Many questions are now in order. Recently the authors of \cite{Hollands:2012sf} proved that the
black strings constructed out with black holes with negative heat capacity
will be perturbatively unstable, proving under certain conditions, the
Gubser-Mitra conjecture \cite{Gubser:2000ec}. It is well known that small black holes in AdS do indeed have negative heat capacity, nevertheless we have explicitly proved the
stability of the black strings for arbitrary radius of the
Schwarzschild-AdS black hole of the transverse section. It would be interesting
to explore how the inclusion of the negative cosmological constant, or the
massless scalar allows to circumvent the hypotheses presented in \cite{Hollands:2012sf}. On the other hand, it is very
likely, that the ground state $AdS_{4}\times R$ obtained by
setting $m=0$ in (\ref{lapse}), and which requires a non-vanishing scalar field, will suffer from the recently discovered non-linear $AdS_{4}$ instability \cite{Bizon:2011gg}. How the scalar and the extended directions produce an imprint in such non-linear process is something that certainly
deserves further exploration.

\subsection*{Acknowledgements}
We thank Christos Charmousis, Gustavo Dotti, Gaston Giribet, Reinaldo Gleiser, Marcela Lagos and Aldo Vera for enlightening comments. A. C work is supported by Fondecyt Grant 11170274 and Proyecto Interno Ucen I+D-2018 CIP 2018020. C. H acknowledges support from Beca CONICYT Doctorado Nacional (2017) 21171394. J. O. work has been funded by Fondecyt Grant 1181047.

\end{document}